\documentclass[twocolumn,prd,showpacs]{revtex4}

\usepackage{amsmath}
\usepackage{amssymb}

\newcommand{\dif}{\mathrm{d}}
\newcommand{\form}[1]{\boldsymbol{\mathrm{#1}}}
\newcommand{\switch}{\leftrightarrow}

\newcommand{\fpder}[2]{\frac{\partial #1}{\partial #2}}

\begin{document}

\title{A general variational principle for spherically symmetric
perturbations in diffeomorphism covariant theories}

\pacs{04.20.Fy, 04.40.Dg}


\author{Michael D. Seifert}
\email{seifert@uchicago.edu}
\author{Robert M. Wald}
\email{rmwa@uchicago.edu}
\affiliation{Enrico Fermi Institute and Department of Physics, University of Chicago, 5640 S. Ellis Ave., Chicago, IL, 60637, USA} 

\begin{abstract}
We present a general method for the analysis of the stability of
static, spherically symmetric solutions to spherically symmetric
perturbations in an arbitrary diffeomorphism covariant Lagrangian
field theory. Our method involves fixing the gauge and solving the
linearized gravitational field equations to eliminate the metric
perturbation variable in terms of the matter variables.  In a wide class of
cases---which include $f(R)$ gravity, the Einstein-{\ae}ther theory of
Jacobson and Mattingly, and Bekenstein's TeVeS theory---the
remaining perturbation equations for the matter fields are second 
order in time. We show how the
symplectic current arising from the original Lagrangian gives rise to
a symmetric bilinear form on the variables of the reduced theory. If
this bilinear form is positive definite, it provides an inner product
that puts the equations of motion of the reduced theory into a
self-adjoint form. A variational principle can then be written down
immediately, from which stability can be tested readily. We illustrate
our method in the case of Einstein's equation with perfect fluid
matter, thereby re-deriving, in a systematic manner, Chandrasekhar's
variational principle for radial oscillations of spherically symmetric
stars. In a subsequent paper, we will apply our analysis to $f(R)$
gravity, the Einstein-{\ae}ther theory, and Bekenstein's TeVeS theory.
\end{abstract}

\maketitle

\section{Introduction}

In recent years, there have been a number of proposals for 
new theories which modify 
general relativity in an attempt to either explain outstanding 
experimental problems or provide a theoretical framework for the analysis 
of new phenomena.  In the first category, we have theories such as 
Carroll {\it et al.}'s $f(R)$ gravity \cite{CDTT}, which attempts to 
modify general relativity to explain the cosmic acceleration, and 
Bekenstein's TeVeS theory \cite{TeVes}, an attempt to formulate a 
covariant version of Milgrom's MOND.  In the second category, we have 
theories such as Jacobson and Mattingly's ``Einstein-{\ae}ther theory'' 
\cite{Aether1}, a toy theory in which Lorentz symmetry is dynamically 
broken by a vector field which is constrained to be unit and timelike.

If such theories are to be phenomenologically viable, they must 
satisfy two criteria, among others.  First, they must possess solutions 
which are quasi-Newtonian, i.e., they must reproduce the dynamics of 
Newtonian gravity up to some small relativistic corrections.  Second, 
these quasi-Newtonian solutions must be stable, or at least must not be 
unstable on time scales sufficiently short to interfere with known 
physical phenomena in Newtonian gravity.  In particular, there must 
exist static, spherically symmetric solutions of the theory, 
corresponding to the interior and exterior of a star, and moreover 
these solutions must not be unstable on a time scale significantly 
shorter than cosmological time scales.

It would therefore be useful to obtain a general method by which the 
stability of an arbitrary Lagrangian field theory might be analyzed.  
The perturbational equations of motion off of some background will be 
a set of linear partial differential equations in terms of some 
perturbational fields. We shall denote these fields $\psi^\alpha$, where 
$\alpha$ is an index running over the collection of perturbational 
fields. Suppose that it were the case that these equations take the 
form
\begin{equation}
	\label{simpleform}
	- \frac{\partial^2}{\partial t^2} \psi^\alpha = 
	{\mathcal{T}^\alpha}_\beta \, \psi^\beta
\end{equation}
where $\mathcal{T}$, the time-evolution operator, is a linear operator 
containing spatial derivatives only. Suppose, further, that we can find 
an inner product $(\cdot, \cdot)$ on the vector space of fields 
$\psi^\alpha$ in which the operator $\mathcal{T}$ is self-adjoint, i.e.,
the domain of $\mathcal{T}$ coincides with the domain of $\mathcal{T}^\dagger$
and
\begin{equation}
	\label{simpleinnerprod}
	(\psi, \mathcal{T} \chi) = (\mathcal{T} \psi, \chi)
\end{equation}
for all $\psi$ and $\chi$ in the domain of $\mathcal{T}$. Let
$\omega_0^2$ denote the greatest lower bound of the spectrum of
$\mathcal{T}$.  Then it can be shown \cite{Waldstability} that if $\omega_0^2 >
0$ then the background solution is stable, whereas if $\omega_0^2 < 0$ then
perturbations exist that grow exponentially on a timescale $\tau =
1/|\omega_0|$. Furthermore, for all $\psi$ in the space of perturbational 
fields, we have (by the Rayleigh-Ritz principle)
\begin{equation}
	\label{simplevarprin}
	\omega_0^2 \leq \langle \mathcal{T} \rangle_\psi = \frac{ (\psi, 
	\mathcal{T} \psi) }{(\psi, \psi)} \, .
\end{equation}
Thus, in practice, then, we can obtain a bound on the spectrum of
$\mathcal{T}$ by plugging various ``trial functions'', $\psi^\alpha$,
into \eqref{simplevarprin}.  If we find a trial function which yields
a negative result, we know that there exist solutions of
\eqref{simpleform} that grow exponentially in time, and we also obtain a
upper bound on the the timescale on which the instability of fastest growth
occurs.

Unfortunately, for a generally covariant theory, the perturbation
equations can never directly arise in the simple form of
\eqref{simpleform}. There will be gauge freedom, so the equations will
not even be deterministic. There will be constraint equations, which
typically will be of lower order in time derivatives than the other
equations. Furthermore, even if one succeeds in suitably fixing a
gauge and solving some subset of the equations so that the remaining
equations take the form of \eqref{simpleform}, there may not exist
an inner product that makes $\mathcal{T}$ self-adjoint, and even if such
an inner product exists, it may be highly nontrivial to find it.

In this paper, we will show that all of the above problems can be
solved in a wide class of diffeomorphism covariant theories in the
context of spherically symmetric perturbations of static, spherically
symmetric solutions. The easiest task is to fix the gauge for such
perturbations, and we make a choice of gauge in Section
\ref{gaugesec}. In Section \ref{constsec}, we analyze the linearized
constraint equations. For spherically symmetric perturbations, there
are two independent constraint equations, namely a ``time-time'' and a
``time-radial'' equation. One of the main results of this paper is our
proof that these two equations can always be reduced to a single
equation of lower differential order. Furthermore, if the original
constraint equations are partial differential equations of second (or
lower) differential order, we show that in a wide class of theories
this resulting single equation 
can be solved algebraically for one of the metric perturbation variables,
thus completely eliminating the constraints.  In addition, we show that
in a wide class of theories, the remaining independent component of the 
gravitational field equations can be used to eliminate the remaining 
metric perturbation variable.  We thereby reduce the theory to solving
a system of partial differential equations for just the matter
perturbation variables.

In the theories that we shall consider here and in a subsequent paper,
the equations of motion of this reduced theory take the form of
\eqref{simpleform}. However, we still have the problem of
determining whether there exists an inner product that makes
$\mathcal{T}$ self-adjoint, and finding this inner product if it does
exist. In this regard, it might be expected that a Hamiltonian
formulation of the theory would be useful, since for a Hamiltonian of
a simple ``kinetic plus potential'' form, the kinetic energy
expression will provide such an inner product. Now, by assumption, the
original theory was obtained from a Lagrangian, and, from this
Lagrangian, one can obtain a Hamiltonian \cite{IyerWald1}. Prior to
gauge fixing and solving the constraints, a Lagrangian and Hamiltonian
for perturbations can be obtained by expanding about the background
solution and keeping the terms quadratic in the perturbation
variables.  However, it is well known that if one substitutes
a choice of gauge into a Lagrangian or Hamiltonian, it no longer
functions as a Lagrangian or Hamiltonian. Similarly, if one
substitutes a solution to some of the field equations (such as the
constraints) into a Lagrangian or Hamiltonian, it no longer functions
as a Lagrangian or Hamiltonian. Thus, it might appear that the
existence of a Lagrangian or Hamiltonian formulation of the original
theory will be useless for obtaining a Lagrangian or Hamiltonian
formulation of the reduced theory, and also useless for finding an
inner product in the reduced theory that makes $\mathcal{T}$
self-adjoint.

In Section \ref{varprinsec}, we make use of the fact that---although
substitution of a gauge choice and/or a solution to some of the field
equations into a Lagrangian or Hamiltonian does not yield a Lagrangian
or Hamiltonian---substitution of a gauge choice and/or a solution to
some of the field equations into the symplectic current still yields a
symplectic current that is conserved when the remaining linearized
equations of motion are satisfied. We show that---in a wide class of
cases---the resulting symplectic current of the reduced theory gives
rise to a bilinear form that makes $\mathcal{T}$ symmetric. If this
bilinear form is positive definite, it provides the desired inner
product, thus enabling us to obtain a variational principle for
determining stability. 

In Section \ref{Chandra} we shall illustrate our method by applying it
to the case of the Einstein-fluid system. We thereby will re-derive in
a systematic fashion the variational principle for analyzing radial
oscillations of spherically symmetric stars in general relativity that
was first obtained by Chandrasekhar \cite{Chandra1,Chandra2}. In a subsequent
paper, the method will be applied to analyze $f(R)$ gravity, the
Einstein-{\ae}ther theory of Jacobson and Mattingly, and to
Bekenstein's TeVeS theory.

Our notation and conventions will generally follow
\cite{Waldgr}. Where relevant, we will use units in which $c = G = 1$.
Boldface symbols will be used to denote differential
forms. For definiteness, we will restrict consideration here to
four-dimensional spacetimes that are spherically symmetric in the sense
of having an $SO(3)$ isometry subgroup whose orbits are
two-spheres. However, all of our analysis and results generalize
straightforwardly to $n$-dimensional spacetimes that are spherically
symmetric in the sense of having an $SO(n-1)$ isometry subgroup whose
orbits are $(n-2)$-spheres.

\section{Lagrangian Formalism \label{lagrangian}}

In this section, we shall briefly review some basic definitions and
constructions in theories derived from a diffeomorphism covariant
Lagrangian. We refer the reader to \cite{LeeWald} and \cite{IyerWald1}
for further details and discussion.

Consider a diffeomorphism covariant Lagrangian four-form
$\form {\mathcal L}$, constructed from dynamical fields $\Psi$
that consist of the spacetime metric and perhaps
additional matter fields, assumed to be described by tensor
fields. For computational convenience, in this paper we will use the
inverse metric $g^{ab}$ rather than the 
metric $g_{ab}$ as the independent dynamical variable describing the
gravitational degrees of freedom.  The first variation of $\form
{\mathcal L}$ can be written in the form 
\begin{equation}
\label{ELeqs}
\delta \form {\mathcal L} = \form {\mathcal E} \delta \Psi
+ \dif \form {\theta} \, ,
\end{equation}
which defines not only the Euler-Lagrange equations of motion,
$\form {\mathcal E} = 0$, but also the symplectic potential
three-form, $\form{  \theta} (\Psi, \delta \Psi)$. In
\eqref{ELeqs}, any tensor indices of $\delta \Psi$ are understood to be
contracted with the corresponding dual tensor indices of
$\form {\mathcal E}$, and, of course, the sum over all fields
comprising $\Psi$ is understood. The antisymmetrized variation of the
symplectic potential yields the symplectic current three-form
$\form{  \omega}(\Psi; \delta_1 \Psi, \delta_1 \Psi)$, defined by
\begin{equation}
\label{omega}
\form{  \omega} = \delta_1 \form {\theta}(\Psi, \delta_2 \Psi)
- \delta_2 \form {\theta}(\Psi, \delta_1 \Psi) \, .
\end{equation}
(Here, $\delta_1$ and $\delta_2$ are taken to be variations along a
two-parameter family, and hence they commute, $\delta_1 \delta_2 =
\delta_2 \delta_1$.)  Applying $\dif$ to this equation and using
\eqref{ELeqs}, we obtain
\begin{equation}
\label{omegacons}
\dif \form{  \omega} = 
- (\delta_1 \form {\mathcal E}) \delta_2 \Psi
+ (\delta_2 \form {\mathcal E}) \delta_1 \Psi \, .
\end{equation}
Thus, the symplectic current is conserved, $\dif \form{  \omega} =
0$, whenever $\delta_1 \Psi$ and $\delta_2 \Psi$ satisfy the
linearized equations of motion.

The Noether current three-form $\form  {J}_\xi$ associated with an
arbitrary vector field $\xi^a$ is defined by
\begin{equation}
\label{J}
\form  {J}_\xi = \form {\theta}(\Psi, \mathcal{L}_\xi \Psi)
- \xi \cdot \form {\mathcal L} \, ,
\end{equation}
where the ``$\cdot$'' denotes the contraction of $\xi^a$ into the first
index of $\form {\mathcal L}$. Applying $\dif$ to this equation, we obtain
\begin{equation}
\label{dJ}
\dif \form  {J}_\xi = - \form {\mathcal E} \mathcal{L}_\xi \Psi \, .
\end{equation}
Thus, $\form{   J}$ is conserved, $\dif \form{   J} = 0$, whenever $\Psi$
satisfies the equations of motion (i.e., $\form {\mathcal E} = 0$) or
whenever $\xi^a$ is a symmetry of $\Psi$ (i.e., $\mathcal{L}_\xi \Psi = 0$).
Finally, we note that a simple calculation \cite{IyerWald1} shows that 
\begin{equation}
\label{deltaJ}
\delta \form{   J}_\xi = -\xi \cdot \form {{\mathcal E}} +
\form {\omega}(\Psi; \delta \Psi, {\mathcal L}_\xi \Psi) + \dif (\xi
\cdot \form {\theta}) \, .
\end{equation}

\section{Gauge fixing \label{gaugesec}}

As noted in the Introduction, generally covariant theories are gauge
theories: they possess ``unphysical degrees of freedom'',
corresponding to diffeomorphisms. In order to obtain deterministic
equations of motion and obtain a variational principle of the sort we
seek, we must eliminate the gauge degrees of freedom. There is no
known algorithm for doing this in general spacetimes. However, in our work,
we will be dealing with spherically symmetric perturbations of static,
spherically symmetric spacetimes, and thus may restrict attention to
spacetimes which are spherically
symmetric but not necessarily static.  For such spacetimes, it is
always possible to put the metric in the following form:
\begin{equation}
  \label{metricform}
  \dif s^2 = - \exp( 2 \Phi(r,t) ) \dif t^2 + \exp (2 \Lambda(r,t))
  \dif r^2 + r^2 \dif \Omega^2
\end{equation}
(See, e.g., \cite{MTW} for the details.)
 Apart from the rotational isometries, the only
diffeomorphisms that preserve the metric form \eqref{metricform} are
redefinitions of the time coordinate of the form $t \rightarrow g(t)$,
for an arbitrary monotonic function $g$. 

For the static background
solution, $\Phi$ and $\Lambda$ can be chosen to be independent of $t$,
in which case the only remaining gauge freedom is $t \rightarrow ct$ for some
constant $c$. Thus, for the background solution, $\Phi$ is unique up
to an additive constant and $\Lambda$ is unique.

Now consider an arbitrary spherically symmetric (but not necessarily
static) perturbation of the background metric. Let $\phi(r,t)$ and
$\lambda(r,t)$ denote, respectively, the perturbations of $\Phi$ and
$\Lambda$.  In other words, $2 e^{-2 \Phi} \phi(r,t)$ is the
perturbation of the $g^{tt}$ metric component, and $-2 e^{-2
\Lambda} \lambda(r,t)$ is the perturbation of the $g^{rr}$
component. The gauge freedom in the perturbed metric is $\delta g^{ab}
\rightarrow \delta g^{ab} + \mathcal{L}_v (g^{(0)})^{ab}$, where 
$(g^{(0)})^{ab}$ is the background metric and $v^a$ is an
arbitrary vector field that generates diffeomorphisms that preserve
the metric form \eqref{metricform}. Since the only such nontrivial
diffeomorphisms are $t \rightarrow g(t)$, the only such nontrivial
$v^a$ is $v^a = h(t) t^a$, where $h(t)$ is a positive function and
$t^a$ is the static Killing field of the background solution. Thus, we
find that $\lambda$ is gauge invariant and $\phi$ has the gauge freedom
\begin{equation}
\label{phifreedom}
\phi(r,t) \rightarrow \phi(r,t) + f(t) \, .
\end{equation}
where $f = dh/dt$ is an arbitrary function of $t$. Thus, our choice of
the metric form \eqref{metricform} eliminates the gauge freedom in the
metric perturbation variables except for the small residual freedom 
\eqref{phifreedom}.

An important consequence of the existence of this residual gauge
freedom is that if $\lambda(r,t)$ and $\phi(r,t)$ along with the
appropriate perturbed matter variables solve the linearized equations
of motion, then if we replace $\phi(r,t)$ by $\phi(r,t) + f(t)$, we
must still obtain a solution to the linearized equations of
motion. This implies that $\phi$ can appear in the equations of motion
only in the form\footnote{If any of the background matter fields are tensor 
fields that have
a nonvanishing time component, then the perturbations of these
components will transform nontrivially, in a manner similar to $\phi$,
under a re-definition of the time coordinate. In that case, some
linear combination(s) of $\phi$ and the matter variables can appear in
undifferentiated form, but, if one treats these combinations as
independent variables, the remaining dependence on $\phi$ can appear
only in the form $\partial \phi / \partial r$ and its derivatives.}
$\partial \phi / \partial r$.

\section{Solving the linearized constraint equations \label{constsec}}

The general analysis of \cite{LeeWald} shows that in any generally
covariant theory, there will be a ``constraint'' on the phase space of
the theory associated with the infinitesimal diffeomorphism generated
by an arbitrary vector field $\xi^a$. In other words, if $\xi^a$ is used
to define the notion of ``time translations'', there will be a
corresponding restriction on the phase space of the theory imposed by
some subset of the equations of motion. If $\xi^a$ is used 
to define the notion of ``time translations'', then, as we shall see
below, the resulting
constraint equations will typically be of lower differential order in
time than the other (so-called ``evolution'') equations of
motion. Consequently, if we wish to get our equations of motion into
the simple form \eqref{simpleform}, it normally will be necessary that
we solve the constraint equations so that---after elimination of
variables---the remaining variables are unconstrained. Remarkably, we
now shall show that in a wide class of diffeomorphism covariant
theories, for spherically symmetric perturbations of static
spherically symmetric spacetimes, the linearized constraint equations
can be solved {\it algebraically} for the metric perturbation
$\lambda$ in terms of the matter perturbation variables. Thus, for
this wide class of theories, the constraints can be easily eliminated.

Our analysis makes use of the fact proved in the Appendix of
\cite{IyerWald2} that for an arbitrary diffeomorphism covariant
theory, the Noether charge two-form $\form{   Q}_\xi$ can always be
defined so that that the Noether current $\form{ J}_\xi$ (as defined 
in \eqref{J}),
associated with an arbitrary vector field $\xi^a$ takes the form
\begin{equation}
\label{constraints}
\form{J}_\xi = \xi^a \form {{\mathcal C}}_a + \dif \form{Q}_\xi
\end{equation}
where, according to the analysis of \cite{LeeWald}, the equations
$\xi^a \form {{\mathcal C}}_a = 0$ are the constraints of the
theory associated with the infinitesimal local symmetry $\xi^a$. (We
will, in effect, rederive \eqref{constraints} by our
calculation leading to \eqref{JV2} below.)  Thus, if we perturb
about a background solution, we obviously obtain
\begin{equation}
\label{deltaconstraints1}
\delta \form{   J}_\xi = \xi^a \delta \form{\mathcal{C}}_a + \dif
(\delta \form{Q}_\xi) 
\end{equation}
On the other hand, we previously noted that $\delta \form{   J}_\xi$ satisfies
\eqref{deltaJ}.
Now suppose that we are perturbing about a solution to the field
equations, $\form {{\mathcal E}} = 0$. Suppose further that this
background solution possesses a Killing field $t^a$ that is also a
symmetry of all of the background matter fields, so that ${\mathcal L}_t
\Psi = 0$. Then, choosing $\xi^a = t^a$, we find from 
\eqref{deltaconstraints1} and \eqref{deltaJ} that
\begin{equation}
\label{deltaconstraints2}
t^a \delta \form {{\mathcal C}}_a = \dif [t \cdot \form {\theta}
  -\delta \form{Q}_t]
\end{equation}
Thus, the linearized constraint equations, $t^a \delta \form {{\mathcal
C}}_a = 0$, associated with $t^a$ are equivalent to the equation
\begin{equation}
\label{dbeta}
\dif \form {\beta} = 0
\end{equation}
where
\begin{equation}
\label{betadef}
\form {\beta} \equiv t \cdot \form {\theta} -\delta \form{   Q}_t \, .
\end{equation}

The replacement of the equation $t^a \delta \form {{\mathcal
C}}_a = 0$ by the equation $\dif \form {\beta} = 0$ need not, in
general, result in any simplification of the equations, i.e., there
may be as many or more independent components of $\form {\beta}$
as $t^a \delta \form {{\mathcal C}}_a$, and the equations $\dif
\form {\beta} = 0$ may be of as high or higher differential order
as the equations $t^a \delta \form {{\mathcal C}}_a =
0$. However, in the case of spherically symmetric perturbations of
static spherically symmetric solutions, the replacement of $t^a \delta
\form {{\mathcal C}}_a = 0$ by $\dif \form {\beta} = 0$ always results
in a major simplification. The reason for this simplification can be
seen as follows. By spherical symmetry, in the coordinates introduced
in the previous section, the three-form $t^a \delta \form {{\mathcal
C}}_a$ must take the form
\begin{equation}
\label{constraintform}
t^a \delta \form {{\mathcal C}}_a = H_1(t,r) \, \dif t \wedge \dif \Omega
+ H_2(t,r) \, \dif r \wedge \dif \Omega
\end{equation}
where $\dif \Omega \equiv \sin \theta \, \dif \theta \wedge \dif \varphi$. Thus, the
constraint equations give rise to two independent equations, namely
$H_1 = 0$ and $H_2 = 0$. By contrast, 
by spherical symmetry, the two-form $\form {\beta}$ must take the form
\begin{equation}
\label{betaform}
\form {\beta} = F(t,r)\, \dif \Omega  \, .
\end{equation}
Thus, $\form {\beta}$ has only one nonvanishing component, and
the equation $\dif \form {\beta} = 0$ then reduces simply to $F =
\rm{const}$. For the situations we shall consider here and in the
subsequent paper, an ``origin'', $r=0$, will be present in the
spacetime.  The two-form $\form {\beta}$ is locally constructed
from the background and perturbed dynamical fields (see
\eqref{betadef}), and thus must be smooth everywhere, including at
$r=0$. However, the spherical volume element $\dif \Omega = \sin \theta
\, \dif \theta \wedge \dif \varphi$ is not smooth at $r=0$.  Consequently, we must have
$F=0$ at $r=0$, and since $F$ is constant, we must have $F=0$
everywhere.  In summary, we have shown that the two independent
components of the linearized constraints take the form
\begin{align}
\label{constraintform2}
H_1(t,r) &= \frac{\partial F}{\partial t} \, , & 
H_2(t,r) &= \frac{\partial F}{\partial r}
\end{align}
and, thus, using the boundary conditions at $r=0$, the two constraint
equations $H_1 = 0$ and $H_2 = 0$ reduce to the single equation, $F =
0$.  Furthermore, it clear from \eqref{constraintform2} that the
equation $F=0$ will be of lower differential order (by one) in the
dynamical variables than the equations $H_1 = 0$ and $H_2 = 0$. Thus,
we obtain a major simplification by replacing the equations $H_1 = 0$
and $H_2 = 0$ by the equation $F=0$.

It remains now only to get an explicit expression for $F$ in an
arbitrary diffeomorphism covariant theory. The calculation of
$\form {\theta}$ is straightforward, and the expression for the
Noether charge $\form  {Q}$ that satisfies \eqref{constraints} can be
determined by following the procedures outlined in the Appendix of
\cite{IyerWald2}. One may then obtain $\form {\beta}$---and,
hence, $F$---from \eqref{betadef}. However, instead of following
this procedure, we shall give a very
simple derivation of a general formula for the
constraints $t^a \form {{\mathcal C}}_a$. By linearizing this
formula, one can then immediately determine $H_1$ and $H_2$. The desired
quantity $F$
can then be determined by inspection from \eqref{constraintform2}.

As previously noted in Section \ref{lagrangian},
for an arbitrary vector field $\xi^a$, we have
\begin{equation}
\label{conserve}
\dif \form{J}_\xi = -\form{\mathcal E} {\mathcal L}_\xi \Psi \, .
\end{equation}
Now, the right side of this equation is an four-form that depends linearly
on $\xi^a$ and contains no higher than first derivatives of $\xi^a$,
i.e., it is of the form
\begin{equation}
\label{xidep}
-\form{ {\mathcal E}} {\mathcal L}_\xi \Psi = (B_a \xi^a 
+ {C^a}_b \nabla_a \xi^b)  \form{  \epsilon} \, ,
\end{equation}
where $\form{  \epsilon}$ is the volume element associated with
the metric and $B_a$ and ${C^a}_b$ are locally constructed out of the
dynamical fields. We can rewrite the right side as
\begin{equation}
\label{UV}
(B_a \xi^a
+ {C^a}_b \nabla_a \xi^b)  \form{  \epsilon} 
= U_a \xi^a \form{  \epsilon} + \dif \form{   V}
\end{equation}
where
\begin{equation}
\label{U}
U_a \equiv B_a - \nabla_b {C^b}_a
\end{equation}
and
\begin{equation}
\label{V}
V_{cde} \equiv {C^a}_b \xi^b \epsilon_{acde}
\end{equation}
Note that $\form V$ does not depend upon derivatives of $\xi^a$. 

Thus, we have proven that for all $\xi^a$ we have
\begin{equation}
\label{JUV}
\dif (\form{ J}_\xi - \form{ V}) = \xi^a U_a \form{  \epsilon} \, .
\end{equation}
However, the only way that this equation can hold for arbitrary
$\xi^a$ is if both sides are
zero. Namely, if $U_a \neq 0$ at some point $p$, then we could find a
smooth $\xi^a$ of compact support such that $U_a \xi^a \geq 0$
everywhere and $U_a \xi^a > 0$ at $p$. However, the integral of the
right side would then be positive, whereas the integral of the left
side vanishes, thereby yielding a contradiction. Thus, we obtain
\begin{equation}
\label{U2}
U_a = B_a - \nabla_b C^b {}_a = 0
\end{equation}
and
\begin{equation}
\label{JV}
\dif (\form{   J}_\xi - \form{   V}) = 0 \, .
\end{equation}
Equation \eqref{U2} is a generalized version of the Bianchi identity,
applicable to an arbitrary diffeomorphism covariant theory, possibly
containing matter fields. Equation \eqref{JV} implies \cite{Waldexact} that
there exists an two-form $\form{   Q}'$ locally constructed out of
$\xi^a$ and the dynamical fields such that
\begin{equation}
\label{JV2}
\form{J}_\xi = \form{   V} + \dif \form{Q}' \, .
\end{equation}
Comparing with \eqref{constraints} and using the fact that $\form{V}$ 
does not depend upon derivatives of $\xi^a$ (the same argument as used 
below \eqref{JUV}), we see that the constraints are given by
\begin{equation}
\label{CV}
\xi^a \form {{\mathcal C}}_a = \form{   V} \, .
\end{equation}

It is worth noting that, taking into account \eqref{V} and 
\eqref{CV}, the constraints take the form
\begin{equation}
\label{consform}
\xi^a \form {{\mathcal C}}_a = C \cdot \form{\epsilon} \, ,
\end{equation}
where $C^a \equiv C^a{}_b \xi^b$, whereas the 
generalized Bianchi identity \eqref{U2} takes the
form
\begin{equation}
\label{Bid}
\nabla_b C^b{}_a = B_a \, .
\end{equation}
We will obtain explicit formulas for $B_a$ and $C^a{}_b$ below.  It
should be emphasized that \eqref{Bid} holds independently of
whether the equations of motion are satisfied.

We wish now to explicitly calculate $\form{V}$. For the purpose of
this calculation, we assume that the Lagrangian $\form {{\mathcal L}}$ 
depends only on the inverse metric $g^{ab}$ and a single matter field
$A^{a_1 a_2 \dots a_n}{}_{b_1 b_2 \dots b_m}$;  the generalization to
the case of more than one matter field is straightforward. We denote
the gravitational equations of motion (obtained by varying
$\form{\mathcal{L}}$ with respect to $g^{ab}$) by
$(\mathcal{E}_G)_{ab} \form{\epsilon}$, and we denote the matter 
equations of motion (obtained by varying $\form{\mathcal{L}}$ with
respect to $A^{a_1 a_2 \dots a_n}{}_{b_1 b_2 \dots b_m}$) by 
$(\mathcal{E}_M)_{a_1 a_2 \dots a_n} {}^{b_1 b_2 \dots b_m}$.
Equation \eqref{conserve} then takes the more explicit form
\begin{multline}
  \label{noetherd}
  \dif \form{J}_\xi = - \left( (\mathcal{E}_G)_{ab} \mathcal{L}_\xi g^{ab} 
  \right. \\
  \left. + (\mathcal{E}_M)_{a_1 a_2 \dots a_n} {}^{b_1 b_2 \dots b_m} 
  \mathcal{L}_\xi A^{a_1 a_2 \dots a_n} {}_{b_1 b_2 \dots b_m} \right)
  \form{\epsilon} 
\end{multline}
Using $\mathcal{L}_\xi g^{ab} = - (\nabla^a \xi^b + \nabla^b \xi^a)$ and
the standard formula for the Lie derivative of tensor field
\begin{multline}
\label{Lie}
\mathcal{L}_\xi A^{a_1 a_2 \dots a_n} {}_{b_1 b_2 \dots b_m} =
\xi^c \nabla_c A^{a_1 a_2 \dots a_n} {}_{b_1 b_2 \dots b_m} \\
{} - \sum_i A^{a_1 \dots c \dots a_n} {}_{b_1 b_2 \dots b_m} \nabla_c
\xi^{a_i} \\
+ \sum_i A^{a_1 a_2 \dots a_n} {}_{b_1 \dots c \dots b_m} \nabla_{b_i} 
\xi^c,
\end{multline}
we can simply read off the formulas
\begin{equation}
\label{B}
B_c = - (\mathcal{E}_M)_{a_1 \dots a_n} {}^{b_1 \dots b_m}
  \nabla_c A^{a_1 \dots a_n} {}_{b_1 \dots b_m} 
\end{equation}
and
\begin{multline}
  \label{C}
  C^c{}_d = 2 g^{ca} (\mathcal{E}_G)_{ad} \\
  - \sum_i A^{a_1
  \dots a_n} {}_{b_1 \dots d \dots b_m} 
  (\mathcal{E}_M)_{a_1 \dots a_n} {}^{b_1 \dots c \dots b_m}\\
{}+ \sum_i
   A^{a_1 \dots c \dots a_n} {}_{b_1 \dots b_m}
  (\mathcal{E}_M)_{a_1 \dots d \dots a_n} {}^{b_1 \dots b_m}.
\end{multline}
In the above equations, the summations run over all possible substitutions 
of the indices $c$ and $d$ into the $i^{\text{th}}$ slot of $A^{a_1 a_2
\dots a_n}{}_{b_1 b_2 \dots b_m}$ and $(\mathcal{E}_M)_{a_1 a_2 \dots a_n} 
{}^{b_1 b_2 \dots b_m}$.  We then see that
\begin{multline}
  \label{udef}
  U_{c} = - 2 \nabla^a (\mathcal{E}_G)_{ac} -
  (\mathcal{E}_M)_{a_1 \dots a_n} {}^{b_1 \dots b_m}
  \nabla_c A^{a_1 \dots a_n} {}_{b_1 \dots b_m} \\ {} - \sum_i
  \nabla_{a_i} \left( (\mathcal{E}_M)_{a_1 \dots c \dots a_n} {}^{b_1
  \dots b_m} A^{a_1 \dots a_n} {}_{b_1 \dots b_m} \right) \\ {} + \sum_i
  \nabla_{b_i} \left( A^{a_1 \dots a_n} {}_{b_1 \dots c \dots b_m}
  (\mathcal{E}_M)_{a_1 \dots a_n} {}^{b_1 \dots b_m}  \right)
\end{multline}
and
\begin{multline}
  \label{vdef}
  V_{def} = \epsilon_{cdef} \Bigg( 2 \xi^a g^{bc} (\mathcal{E}_G)_{ab} \\
  + \sum_i \xi^{b_i} 
  A^{a_1 \dots a_n} {}_{b_1 \dots b_m}
  (\mathcal{E}_M)_{a_1 \dots a_n} {}^{b_1 \dots c \dots b_m} \\
  \left. {} - \sum_i
  \xi^{a_i} A^{a_1 \dots c \dots a_n} {}_{b_1 \dots b_m}
  (\mathcal{E}_M)_{a_1 \dots a_n} {}^{b_1 \dots b_m}  \right)
\end{multline}
Equation \eqref{vdef} is our desired formula
for the constraints, which can be readily evaluated in cases of
interest. 

\section{Eliminating the Metric Perturbation Variables \label{metpertsec}}

We now wish to solve the linearized field equations
\begin{equation}
\label{lineqs}
(\delta \mathcal{E}_G)_{ab} = 0\, , \,\,\,\, \delta \mathcal{E}_M = 0
\end{equation}
for spherically symmetric perturbations of a static, spherically
symmetric background solution. We shall show that, in a wide class of
cases, it is possible to solve these equations {\it algebraically} for
the metric perturbation variables, thereby reducing the problem to
solving the linearized matter equations $\delta \mathcal{E}_M = 0$ for
the matter variables alone.

To begin, we note that from the form of $C_{ab}$ (see \eqref{C}),
together with the fact that the background equations of motion 
$(\mathcal{E}_G)_{ab} = 0$, $\mathcal{E}_M = 0$ are
satisfied, it follows immediately that \eqref{lineqs} is equivalent to
\begin{equation}
\label{lineqs2}
\delta C_{ab} = 0\, , \,\,\,\, \delta \mathcal{E}_M = 0 \, .
\end{equation}
Next, we note that by spherical symmetry, there are only four
independent components of
$\delta C_{ab}$, which can be chosen to be the $tt$,
$tr$, $rr$, and $\theta \theta$ components. 
However, we showed in the previous section that the $tt$
and $tr$ components of these equations can be
replaced by a single equation of lower differential order, namely
$F=0$. Furthermore, the linearization of the
generalized Bianchi identity \eqref{Bid} off of a background solution 
(where $C^a {}_b = 0$) yields
\begin{equation}
\label{Bid2}
\nabla_a (\delta C^a{}_b) = \delta B_b \, .
\end{equation}
If we impose the linearized
matter equations of motion, $\delta \mathcal{E}_M = 0$, then
$\delta B_b = 0$. This implies that if
the $tt$, $tr$, and $rr$ components of $\delta C_{ab} = 0$ are satisfied,
and if the linearized matter equations of motion, $\delta \mathcal{E}_M = 0$,
hold, then the $\theta \theta$ component
of $\delta C_{ab} = 0$ is automatically satisfied.
Thus, the full set of linearized equations \eqref{lineqs}
may be replaced by
\begin{equation}
\label{lineqs3}
F= 0 \,, \,\,\,\, (\delta C)_{rr} = 0\, , \,\,\,\, 
\delta \mathcal{E}_M = 0
\end{equation}

We now restrict consideration to theories in which the field equations
are at most second order in derivatives of the metric, and such that
the second derivatives of the metric appear only in the form of
curvature. This is the case for general relativity with typical forms
of matter, as well as for Bekenstein's
TeVeS theory. It is also the case for $f(R)$ gravity after a suitable
redefinition of variables, since that theory can be recast as a
scalar-tensor theory. 

We now analyze the possible dependence of $\delta C_{tt}$ on $\lambda$
in order to determine the possible dependence of $F$ on this
quantity. By the Bianchi identity, the $tt$- and $tr$-components of the
equations $\delta C_{ab} = 0$ cannot contain second time derivatives
of the metric perturbation quantities $\phi$ and $\lambda$, since
otherwise the left side of \eqref{Bid2} would contain third time
derivatives of the metric perturbation quantities, which could not be
canceled by terms on the right side, which, by assumption, contain at
most second derivatives (see \eqref{B}). (Recall that the
generalized Bianchi identity \eqref{Bid} holds independently of the
field equations, so its linearization \eqref{Bid2} off of a solution
must hold if $\phi$ and $\lambda$ are taken to be arbitrary functions
of $r$ and $t$.) Thus, in particular, the expression $\delta C_{tt}$
cannot contain a term in $\partial^2 \lambda/ \partial t^2$. On the
other hand, for a diagonal metric like \eqref{metricform}, for
each coordinate $x^\mu$, the Riemann tensor cannot contain a term of
the form $\partial^2 g^{\mu \mu}/\partial x^{\mu 2}$, since the
formula for the Riemann tensor involves antisymmetrizations over
components. Choosing $x^\mu = r$, we see that the Riemann tensor
cannot depend on $\partial^2 g^{rr}/\partial r^2$. This implies that
the expression $\delta C_{tt}$ cannot contain a term of the form
$\partial^2 \lambda / \partial r^2$. Finally, since the background
spacetime is static and thus invariant under time reflection $t
\rightarrow -t$, it follows that the expression $\delta C_{tt}$ must
not change sign under time reversal and thus cannot contain a term of
the form $\partial \lambda / \partial t$ or $\partial^2 \lambda /
\partial r \partial t$. Thus, we conclude that for theories in which
the field equations are at most second order in derivatives of the
metric, and are such that the second derivatives of the metric appear
only in the form of curvature, the quantity $\lambda$ can appear in
the expression $\delta C_{tt}$ only in the form of $\lambda$ and
$\partial \lambda / \partial r$. However, as noted previously, the
expression $F$ must be of lower differential order (by one) than
$\delta C_{tt}$ in all variables. Consequently, $F$ can depend only
algebraically on $\lambda$.

Next, we analyze the possible dependence of $\delta C_{tr}$ on $\phi$
in order to determine the possible dependence of $F$ on this
quantity. As noted at the end of Section \ref{gaugesec}, $\phi$ can appear in the
linearized field equations only in the form of $\partial \phi /
\partial r$ and its derivatives.  However, since $\delta C_{tr}$ must
be odd under time reflection, $\phi$ can appear in $\delta C_{tr}$
only in the form of $\partial \phi / \partial t$ and its
$r$-derivatives. Since, by assumption, the field equations contain no
higher than second derivatives of the metric perturbation quantities,
it follows that $\phi$ can appear in $\delta C_{tr}$ only in the form
of $\partial^2 \phi / \partial r \partial t$. Suppose now that $\delta
C_{tr}$ contained a nonvanishing term proportional to $\partial^2 \phi
/ \partial r \partial t$. Then, the $r$-component of the Bianchi
identity \eqref{Bid} would contain a term proportional to $\partial^3
\phi / \partial r \partial t^2$.  The only other term in the
Bianchi identity \eqref{Bid} that can contain third derivatives of the
metric perturbation quantities is $\partial (\delta C^r{}_r ) / \partial
r$. In order to cancel the term proportional to $\partial^3 \phi /
\partial r \partial t^2$, it is necessary that $\delta C_{rr}$ contain
a term proportional to $\partial^2 \phi / \partial t^2$. However, this
is impossible, since the field equations can contain $\phi$ only in
the form of $\partial \phi / \partial r$ and its derivatives.
Consequently, $\delta C_{tr}$ cannot have any dependence whatsoever
upon $\phi$. It then follows immediately that $F$ cannot depend upon
$\phi$.

Putting together the results of the previous two paragraphs, we
conclude that for theories in which the field equations are at most
second order in derivatives of the metric, and are such that the
second derivatives of the metric appear only in the form of curvature,
{\it we can solve the equation $F=0$ algebraically for $\lambda$ in
terms of the matter variables.}

Next, we consider the equation $(\delta C)_{rr} = 0$. For exactly the
same reason as $(\delta C)_{tt}$ cannot contain second time derivatives
of the metric perturbation quantities, it follows that $(\delta
C)_{rr}$ cannot contain second derivatives of these quantities with
respect to $r$, and thus cannot contain a term of the form $\partial^2
\phi / \partial r^2$. By time reflection symmetry, it also cannot
contain a term of the form $\partial \phi/\partial t$ or $\partial^2
\phi / \partial r \partial t$. However, as already noted above, $\phi$
can appear in the linearized field equations only in the form of
$\partial \phi / \partial r$ and its derivatives. Thus, we also cannot
have a term in $(\delta C)_{rr}$ of the form $\partial^2 \phi /
\partial t^2$ (as we already noted above) or
$\phi$. Consequently, $(\delta C)_{rr}$ can depend only algebraically
on $\partial \phi / \partial r$. Thus, since we have already
eliminated $\lambda$ in terms of the matter variables {\it for the
class of theories considered here, we can solve the equation $(\delta
C)_{rr}= 0$ algebraically for $\partial \phi / \partial r$ in terms of
the matter variables.}

In summary, we have seen that we can eliminate the metric perturbation
variables algebraically by solving the equation $F=0$ for $\lambda$
and solving the equation $(\delta C)_{rr}= 0$ for $\partial \phi /
\partial r$. We may then substitute these solutions into the
linearized matter equations
\begin{equation}
\label{lineqs4}
\delta \mathcal{E}_M = 0
\end{equation}
to obtain a system of equations involving only the unknown matter
variables. The perturbation problem has thus been reduced to solving
these equations.

\section{Obtaining a Variational Principle \label{varprinsec}}

Having fixed the gauge and completely eliminated the remaining metric
perturbation variables by the procedures of the previous three
sections, we may find that the system of equations \eqref{lineqs4} for
the matter variables can be put in the general form
\eqref{simpleform}. Indeed, we shall see in the next section and in a
subsequent paper that this is the case in a wide class of
theories. Nevertheless, even if the reduced equations take the form
\eqref{simpleform}, our work is not completed because it is not
straightforward to determine if there exists an inner product that
satisfies \eqref{simpleinnerprod}, and---even if such an inner product
does exist---it is not straightforward to find it explicitly. This is
particularly true if \eqref{simpleform} is a complicated system of
equations. In the absence of an inner product that makes
$\mathcal{T}^\alpha{}_\beta$ self-adjoint, we will not be able to
formulate a variational principle, and it will not be straightforward
to analyze stability.

We now shall show that the fact that the original theory was derived
from a Lagrangian will enable us---at least in a very wide class of
cases---to determine whether the desired inner product exists and to find
it explicitly if it does. When the desired inner product does exist, we
will thereby be able to immediately
write down the variational principle \eqref{simplevarprin} for
determining stability.

At first sight, it might appear that the fact that the original theory
was derived from a Lagrangian would be of little use. It is true that
one can obtain a Lagrangian for the linearized theory by expanding the
Lagrangian of the exact theory to quadratic order about the background
solution. However, if one substitutes the gauge choice made in Section
\ref{gaugesec} into this Lagrangian, the resulting object no longer functions as a
Lagrangian. Similarly, even if one found a Lagrangian for the gauge-fixed
theory, when one substitutes the solution for the metric perturbation 
variables found in Section \ref{metpertsec} into this
Lagrangian, the resulting object would again fail
to function as a Lagrangian. Thus, it is far from obvious how to
obtain a Lagrangian for the reduced theory.

Nevertheless, the fact that the original theory was derived from a
Lagrangian leaves an important imprint on the reduced theory. 
As discussed in Section \ref{lagrangian}, in the
original, exact theory, one can define a symplectic current three-form
$\form{  \omega} (\Psi; \delta_1 \Psi, \delta_2 \Psi_2)$, which
is constructed out of a background solution $\Psi$ and two linearized
perturbations, $\delta_1 \Psi$ and $\delta_2 \Psi$, such that
$\form{  \omega}$ depends linearly on $\delta_1 \Psi$ and $\delta_2
\Psi$. Furthermore, the symplectic current satisfies the property that it is
conserved, i.e.,
\begin{equation}
\label{domega}
\dif \form{\omega} = 0
\end{equation}
whenever $\delta_1 \Psi$ and $\delta_2 \Psi$ satisfy the linearized
equations of motion. Substitution of gauge choices for $\delta_1 \Psi$
and $\delta_2 \Psi$ and/or elimination of variables via some of the 
linearized field equations will not affect the conservation of 
$\form{\omega}$. Thus, we automatically obtain a conserved symplectic
current for the reduced theory.

The conditions arising from the existence of a conserved symplectic
current $\form {\omega}$ are most conveniently formulated in
terms of the pullback, $\bar{\form{\omega}}$, of
$\form {\omega}$ to the static hypersurfaces of the background
solution, i.e., the hypersurfaces orthogonal to the static Killing
field $t^a$. Since $\dif \form {\omega}$ is a four-form in a 
four-dimensional space, the
condition that $\dif \form {\omega} = 0$ is equivalent to $t \cdot \dif
\form {\omega} = 0$. By a standard identity, we have
\begin{equation}
\label{Lieomega}
t \cdot \dif \form {\omega} = {\mathcal L}_t \form {\omega} - 
\dif (t \cdot \form {\omega}) \, .
\end{equation}
Thus, when the equations of motion hold, we have
\begin{equation}
\label{Lieomega2}
{\mathcal L}_t \form {\omega} = \dif (t \cdot \form {\omega}) \, .
\end{equation}
Pulling this equation back to the static hypersurfaces, we obtain
\begin{equation}
\label{Lieomega3}
{\mathcal L}_t \bar{\form{\omega}} 
= \dif (\overline {t \cdot \form {\omega}}) \, .
\end{equation}

In the theories we shall consider here and in a subsequent paper, the
pullback of the symplectic current for the reduced theory takes the form
\begin{equation}
\label{baromegaform}
\bar{\form{\omega}} = \form{W}_{\alpha \beta}
\left( \fpder{\psi_1^\alpha}{t} \psi_2^\beta -
\fpder{\psi_2^\alpha}{t} \psi_1^\beta \right)
\end{equation}
where $\psi^\alpha$ denotes the dynamical variables for the reduced
theory.  Here, the three-form
$\form  {W}_{\alpha \beta}$ is constructed from the quantities appearing
in the background solution, and thus is independent of $t$, i.e.,
${\mathcal L}_t \form  {W}_{\alpha \beta} = 0$. Thus, we obtain
\begin{multline}
\label{Lieomega4}
{\mathcal L}_t \bar{\form{\omega}} = \form  {W}_{\alpha \beta}
\left( \frac{\partial \psi_1^\alpha}{\partial t} \frac{\partial
  \psi_2^\beta}{\partial t} - 
\frac{\partial \psi_2^\alpha}{\partial t} \frac{\partial
  \psi_1^\beta}{\partial t} \right. \\
\left.
+ \psi_1^\alpha \frac{\partial^2 \psi_2^\beta}{\partial t^2} -
\psi_2^\alpha \frac{\partial^2 \psi_1^\beta}{\partial t^2} \right) \, .
\end{multline}
However, the linearized equations of motion \eqref{simpleform} hold if
and only if $\partial^2 \psi^\alpha / \partial t^2 =
\mathcal{T}^\alpha{}_\beta \psi^\beta$. 
Thus, we find that the quantity
\begin{equation}
\label{W}
\form{W}_{\alpha \beta}
\left( \frac{\partial \psi_1^\alpha}{\partial t} \frac{\partial
  \psi_2^\beta}{\partial t} - 
\frac{\partial \psi_2^\alpha}{\partial t} \frac{\partial
  \psi_1^\beta}{\partial t}
+ \psi_1^\alpha \mathcal{T}^\beta {}_\gamma \psi_2^\gamma 
- \psi_2^\alpha \mathcal{T}^\beta {}_\gamma \psi_1^\gamma \right)
\end{equation}
must be an exact form on any static hypersurface, $\Sigma_t$. Now,
$\psi_1^\alpha$, $\partial \psi_1^\alpha / \partial t$, $\psi_2^\alpha$, and
$\partial \psi_2^\alpha / \partial t$ are all freely specifiable
initial data at $t=0$ for 
the system of equations \eqref{simpleform}. Thus, if we
choose
$\psi_1^\alpha$, $\partial \psi_1^\alpha / \partial t$, $\psi_2^\alpha$, and
$\partial \psi_2^\alpha / \partial t$
to be arbitrary smooth functions of compact
support (or of sufficiently rapid decay), the integral of \eqref{W}
over $\Sigma_0$ must vanish.  Inspecting the first two terms of \eqref{W},
we see that $\form{W}_{\alpha \beta}$ must be symmetric in its ``field 
space indices'', i.e.,
\begin{equation}
\label{W2}
\form  {W}_{\alpha \beta} = \form  {W}_{\beta \alpha}
\end{equation}
and inspecting the last two terms of \eqref{W}, we see that for all
$\psi_1^\alpha$ and $\psi_2^\alpha$ of compact support (or of
sufficiently rapid decay), we have 
\begin{equation}
\label{W3}
\int_{\Sigma_0} \form  {W}_{\alpha \beta} (\psi_1^\alpha
\mathcal{T}^\beta {}_\gamma \psi_2^\gamma - \psi_2^\alpha
\mathcal{T}^\beta {}_\gamma \psi_1^\gamma) = 0.
\end{equation}

Now, suppose that $\form  {W}_{\alpha \beta}$ is positive definite in the sense
that 
\begin{equation}
\label{W4}
\int_{\Sigma_0} \form  {W}_{\alpha \beta} \psi^\alpha \psi^\beta > 0
\end{equation}
for all $\psi^\alpha \neq 0$. Then
\begin{equation}
\label{innerprod}
(\psi_1, \psi_2) \equiv \int_{\Sigma_0} \form  {W}_{\alpha \beta}
\psi_1^\alpha \psi_2^\beta 
\end{equation}
defines an inner product. Equation \eqref{W3} is then precisely the
statement that the operator $\mathcal{T}^\alpha{}_\beta$ is symmetric
in this inner product. We thereby obtain a variational
principle\footnote{To obtain a variational principle, we actually require
$\mathcal{T}^\alpha{}_\beta$ to be self-adjoint, not merely
symmetric. However, since $\mathcal{T}^\alpha{}_\beta$ is real, it
always admits self-adjoint extensions. If $\mathcal{T}^\alpha{}_\beta$
admits more than one self-adjoint extension (i.e., if
$\mathcal{T}^\alpha{}_\beta$ is not essentially self-adjoint on the
initial domain of smooth functions of compact support), this simply
means that additional boundary conditions must be supplied in order to
make the dynamics specified by \eqref{simpleform} well defined.},
as desired.

On the other hand, if $\form  {W}_{\alpha \beta}$ fails to be positive
definite (or negative definite), then there does not appear to be any
reason to expect that an inner product exists that makes
$\mathcal{T}^\alpha{}_\beta$ self-adjoint\footnote{If $\form  {W}_{\alpha
\beta}$ fails to be positive definite but is non-degenerate, then we
obtain a Kre\u{\i}n space, in which $\mathcal{T}^\alpha{}_\beta$ is
symmetric. However, this does not provide us with a variational
principle to determine stability.}. In that case, we do not expect
that there is a variational principle to determine stability, and one
presumably must work directly with the equations of motion to analyze
stability.

Finally, it is worth noting that if $\form  {W}_{\alpha \beta}$ is
positive definite, then given that $\mathcal{T}^\alpha{}_\beta$ must
satisfy \eqref{W3}, it is easily verified that
\begin{equation}
\label{hamiltonian}
h = \frac{1}{2} \int_{\Sigma_0} \form  {W}_{\alpha \beta}
\left(\frac{\partial \psi^\alpha}{\partial t} \frac{\partial
  \psi^\beta}{\partial t}
+ \psi^\alpha \mathcal{T}^\beta{}_\gamma \psi^\gamma \right)
\end{equation}
defines a Hamiltonian for the dynamics given by \eqref{simpleform}
associated with the symplectic form 
\begin{equation}
\label{sympform}
\Omega(\psi_1, \psi_2) = \int_{\Sigma_0} \form {\omega} (\Psi; \psi_1,
\psi_2) \, .
\end{equation}

\section{Illustration: Chandrasekhar's Variational Principle}
\label{Chandra}
As a concrete example of the procedure outlined here, let us consider
Einstein gravity minimally coupled to an isentropic perfect fluid.  A
variational principle for the radial oscillations of static,
spherically symmetric stars was derived by Chandrasekhar
\cite{Chandra1, Chandra2} by a direct analysis of the equations of
motion. We now shall re-derive this variational principle in a much
more systematic and direct way using the method described in this
paper.

In order to apply our method, it is essential that the equations of
motion be derived from a Lagrangian. 
For general matter minimally coupled to $g^{ab}$, the Lagrangian will
be of the form
\begin{equation}
\form{\mathcal{L}} = \frac{1}{16 \pi}R \form{\epsilon} +
\form{\mathcal{L}}_{\text{mat}}[\Psi, g^{ab}]
\end{equation}
where $A$ denotes the collection of matter fields, with tensor indices
suppressed.  For the case of a
perfect fluid, there have been there are several approaches that have
been used to provide a Lagrangian formulation; see \cite{Brown} for an
overview. We will use the ``Lagrangian coordinate'' method. In this
formalism, we introduce an abstract three-dimensional manifold,
$\mathcal M$, of fluid worldlines, equipped with a volume three-form
$\bf{N}$.  The fluid is then described by a map $\chi: M \rightarrow
\mathcal{M}$, which assigns to each $x$ in the spacetime manifold $M$
the fluid worldline that passes through $x$.  By introducing
coordinates, $X^A$, on $\mathcal 
M$, where $A = 1, 2, 3$, we can represent $\chi$ by the three scalar functions
$X^A(x)$ on $M$, which we view as the independent dynamical variables
of the fluid.

We define a three-form $N_{abc}$ on spacetime---representing the
``density of fluid worldlines'' or the ``particle number
density''---by
\begin{equation}
\label{Ndef}
N_{abc} = N_{ABC}(X) \nabla_a X^A \nabla_b X^B \nabla_c X^C \, .
\end{equation}
In terms of $N_{abc}$ we define the scalar particle number density
$\nu$ by 
\begin{equation}
\label{nuandN}
\nu^2 = \frac{1}{6} N_{abc} N^{abc},
\end{equation}
and we define the fluid four-velocity $U^a$ by
\begin{equation}
\label{NvsU}
N_{abc} = \nu \epsilon_{abcd} U^d \, ,
\end{equation}
where $\epsilon_{abcd}$ is the spacetime volume four-form.  The
Lagrangian four-form for the perfect fluid is then given by
\begin{equation}
\form{\mathcal{L}}_\text{mat}  = - \varrho (\nu)
\form{\epsilon}, 
\end{equation}
where $\varrho(\nu)$ is an arbitrary function of the comoving particle
density $\nu$. The choice of the function
$\varrho(\nu)$ corresponds to the choice
of equation of state of the fluid (see \eqref{rhoandP} below).

We will need both the equations of motion and the formula for
$\form{\omega}$ in our analysis, so we now proceed to derive these.
As usual, the ``gravitational part'' of the Lagrangian,
$\form{\mathcal{L}}_G = (1/16\pi) R \form{\epsilon}$,
contributes $(1/16 \pi) G_{ab}$ to the gravitational equations of
motion and contributes \cite{BurnettWald} 
\begin{equation}
\label{omegagrav}
\omega_\text{grav}^a = S^{a} {}_{bc} {}^d {}_{ef}
(\delta_2 g^{bc} \nabla_d \delta_1 g^{ef} - \delta_1 g^{bc}
\nabla_d \delta_2 g^{ef})\, ,
\end{equation}
to the symplectic current, where 
\begin{multline}
S^{a} {}_{bc} {}^d {}_{ef} = \frac{1}{16 \pi} \left( 
  \delta^a {}_e \delta^d {}_c g_{bf} 
  - \frac{1}{2} g^{ad} g_{be} g_{cf} 
  - \frac{1}{2} \delta^a {}_b \delta^d {}_c g_{ef} \right. \\
  \left. - \frac{1}{2} \delta^a {}_e \delta^d {}_f g_{bc} 
  + \frac{1}{2} g^{ad} g_{bc} g_{ef} \right)
\end{multline}
and we have defined $\omega^a$, the dual to $\form{\omega}$, such that 
\begin{equation}
  \label{omegaadef}
  \omega_{bcd} = \omega^a \epsilon_{abcd}
\end{equation}
with $\omega_{bcd}$ defined by \eqref{omega}.

To obtain the matter equations of motion and the contribution of the
matter fields to to $\omega^a$, we vary $\form{\mathcal{L}}_\text{mat}$
with respect to the dynamical fields $X^A$.
The variation of the number density 
three-form is given by (see \cite{Brown})
\begin{equation}
  \delta N_{abc} = 3 \nabla_{[a} (N_{ABC} \delta X^A) \nabla_b X^B
  \nabla_{c]} X^C. 
\end{equation}
(The antisymmetrization in this expression is, of course, over the
spacetime indices only, not the fluid-space indices.
Note that the fluid-space coordinates $X^A$ are simply scalars as far
as the spacetime derivative operator $\nabla_a$ is concerned, and thus
no metric variations, $\delta g^{ab}$, appear in this expression.  The
antisymmetrization in this expression)
Using \eqref{nuandN} and $\delta \nu = \delta(\nu^2) / 2 \nu$, we
obtain 
\begin{multline}
\label{mattervar}
\delta (- \varrho \form{\epsilon}) =
\left[ - \frac{\varrho'}{2 \nu} \nabla_a (N_{ABC} 
\delta X^A) \nabla_b X^B \nabla_c X^C N^{abc} \right. \\
\left. - \frac{1}{2} \left(  
\frac{\varrho'}{2 \nu} N_{acd} N_{b} {}^{cd} - \varrho 
g_{ab} \right) \delta g^{ab} \right] \form{\epsilon} \, ,
\end{multline}
where $\varrho'$ denotes the derivative of $\varrho(\nu)$ with respect
to $\nu$. The second term is brackets is the functional derivative of
$\form{\mathcal{L}}_\text{mat}$ with respect to $g^{ab}$ and thus
is equal to $-\frac{1}{2}$ times the matter stress-energy tensor, $T_{ab}$;
this quantity 
provides the contribution of $\form{\mathcal{L}}_\text{mat}$ to
the gravitational field equations. Using \eqref{NvsU} and
the identity $\epsilon_{abef} \epsilon^{cdef} = - 4 \delta^{[c} {}_a
\delta^{d]} {}_b$, we obtain
\begin{equation}
N_{acd} N_{b} {}^{cd} = 2 \nu^2 (g_{ab} + U_a U_b),
\end{equation}
and thus the stress-energy tensor can be rewritten in more familiar
terms as 
\begin{equation}
T_{ab} = \varrho' \nu U_a U_b + (\varrho' \nu - 
\varrho ) g_{ab}.
\end{equation}
This can be recognized as the standard stress-energy for a perfect fluid
\begin{equation}
T_{ab} = (\rho + P) U_a U_b + P g_{ab}
\end{equation}
under the identifications
\begin{align}
\label{rhoandP}
\varrho &\to \rho & \varrho' \nu - \varrho &\to P.
\end{align}

Integrating the first term in \eqref{mattervar} by parts, we obtain
\begin{multline}
\label{matterEOMs}
-\frac{\varrho'}{2 \nu} \nabla_a (N_{ABC} \delta X^A) \nabla_b 
X^B \nabla_c X^C N^{abc} \\ 
= \nabla_a \left( -\frac{\varrho'}{2 \nu} 
N_{ABC} \delta X^A \nabla_b X^B \nabla_c X^C N^{abc} \right) \\
+ N_{ABC} 
\delta X^A \nabla_a \left( \frac{\varrho'}{2 \nu} \nabla_b 
X^B \nabla_c X^C N^{abc} \right)
\end{multline}
From the first term on the right-hand side, we can read off the
presymplectic current $\theta^a_\text{mat}$:
\begin{equation}
\theta^a_\text{mat} = - \frac{\varrho'}{2 \nu} N_{ABC} \delta
X^A \nabla_b X^B \nabla_c X^C N^{abc}
\end{equation}
We can then take the antisymmetrized second variation of
$\form{\theta}_\text{mat} = \theta \cdot \form{\epsilon}$, as in
\eqref{omega}, to obtain an expression for the symplectic current
three-form $\form{\omega}_\text{mat}$. By a straightforward calculation,
we obtain
\begin{multline}
\label{omegamat}
\omega^a_\text{mat} = \left[ \delta_1 g^{bc} K^a {}_{bc \, A} +
  \delta_1 X^B L^a {}_{AB} \right. \\  
  \left. + \nabla_b \delta_1 X^B M^{ab} {}_{AB} \right] \delta_2 X^A 
  - [1 \switch 2] \, ,
\end{multline}
with $\omega^a_\text{mat}$ defined analogously to \eqref{omegaadef}
and the tensors $K^a {}_{bc \, A}$, $L^a {}_{AB}$, and $M^{ab}
{}_{AB}$ defined by
\begin{widetext}
\begin{multline}
K^a {}_{bc \, A} = - \frac{\varrho'}{2 \nu} \bigg( - \frac{1}{2} g_{bc}
N_{ADE} \nabla_d X^D \nabla_e X^E N^{ade}
+ 2 N_{ABD} \nabla_b X^B \nabla_d X^D N^a {}_c {}^d
+ \delta^a {}_b N_c {}^{de} N_{ADE} \nabla_d X^D \nabla_e X^E \bigg) \\  
+ \frac{1}{8 \nu^3} (\varrho'' \nu - \varrho') N_{bde} N_c {}^{de}
N_{AFG} \nabla_f X^F \nabla_g X^G N^{afg},
\end{multline}
\begin{multline}
\label{Ldef}
L^a {}_{AB} = - \frac{\varrho'}{2 \nu} \left( \partial_B N_{ACD} \nabla_c X^C 
\nabla_d X^D N^{acd} + 3 N_{ACD} \nabla_c X^C \nabla_d X^D \partial_E N_{BFG}
\nabla^{[a} X^E \nabla^c X^F \nabla^{d]} X^G \right) \\ 
+ \frac{1}{4 \nu^3} (\varrho'' \nu - \varrho') N_{ACD} \nabla_b X^C
\nabla_c X^D N^{abc} \partial_E N_{BFG} \nabla_e X^E \nabla_f
X^F \nabla_g X^G N^{efg}, 
\end{multline}
and
\begin{multline}
\label{Mdef}
M^{ab} {}_{AB} = - \frac{\varrho'}{2 \nu} \left( 2 N_{ABC} \nabla_c X^C
N^{abc} + 3 N_{ACD} \nabla_c X^C \nabla_d X^D N_{BEF} g^{b[a}
\nabla^c X^E \nabla^{d]} X^F \right) 
\\ {} + \frac{1}{4 \nu^3} (\varrho'' \nu - \varrho') (N_{ACD}
\nabla_c X^C \nabla_d X^D N^{acd}) (N_{BEF} \nabla_e X^E \nabla_f X^F
N^{bef}) \, .
\end{multline}
\end{widetext}
(The antisymmetrizations in \eqref{Ldef} and \eqref{Mdef} are again
over tensor indices only.)

The second term in \eqref{matterEOMs} yields the matter equations of 
motion.
Since the Lagrangian coordinates $X^A$ are scalars, we have $\nabla_{[a} 
\nabla_{b]} X^A = 0$, and since $N^{abc}$ is completely antisymmetric,
the vanishing of the second term for arbitrary $\delta X^A$ is equivalent to  
\begin{multline}
\label{feom}
0 = \nabla_a \left( \frac{\varrho'}{2 \nu} \nabla_b 
X^B \nabla_c X^C N^{abc} \right) \\ = \frac{1}{2}
\nabla_b 
X^B \nabla_c X^C \nabla_a \left( \frac{\varrho'}{\nu} N^{abc} \right) \, .
\end{multline}
This can be put in a more recognizable form by rewriting it in terms
of the four-velocity $U^a$
\begin{equation}
\label{euler}
0 = \frac{1}{2} \nabla_b 
X^B \nabla_c X^C \nabla_{[a} \left( \varrho' U_{d]} \right) 
\epsilon^{abcd} \, .
\end{equation}
This, in turn, is equivalent to 
$U^a \nabla_{[a} ( \varrho' U_{d]} ) = 0$, which is
just the relativistic Euler equation \cite{Brown}.

We now restrict our attention to spherically symmetric perturbations
of static, spherically symmetric solutions. In order to obtain our
desired variational principle, we must do the following: 
\begin{enumerate}
\item Define our choice of variable(s) to describe the fluid
  perturbations;  
\item Obtain an expression for $F$ and solve the equation $F=0$
  algebraically for $\lambda$; 
\item Write down the equation $\delta C_{rr} = 0$ and solve this
  equation algebraically for $\partial \phi/\partial r$; 
\item Substitute the solutions for $\lambda$ and $\partial
  \phi/\partial r$ into the linearization of the matter equations of
  motion \eqref{feom}, thereby rewriting this equation purely in terms
  of the perturbed fluid variable(s); 
\item Determine if this equation takes the form \eqref{simpleform}
  and, if so, read off the operator $\mathcal{T}^\alpha{}_\beta$; 
\item Evaluate the pullback of $\form{\omega}$---or, equivalently, the
  time component of $\omega^a$;
\item Determine if $\form{\omega}$ takes the form
  \eqref{baromegaform}, and, if so, read off $\form{W}_{\alpha
    \beta}$; and  
\item Determine if $\form{W}_{\alpha \beta}$ defines a positive
  definite inner product and, if so, write down the variational
  principle \eqref{simplevarprin}.  
\end{enumerate}
Although each of these steps may require some
algebra, we emphasize that none of the steps require any ingenuity,
and we are guaranteed to succeed in obtaining a variational principle
unless the matter equations of motion obtained in step (iv) fail to 
take the form \eqref{simpleform}, the pullback of $\form{\omega}$
fails to take the form \eqref{baromegaform}, or $\form{W}_{\alpha
  \beta}$ fails to define a positive definite inner product.

To describe the fluid perturbations, we choose the coordinates $X^A
\equiv \{X^R, X^\Theta, X^\Phi \}$ on fluid space so that in the
static background solution we have $X^R =r$, $X^\Theta = \theta$, and
$X^\Phi = \varphi$. Since we consider only spherically symmetric
perturbations, we have $\delta X^\Theta = 0$ and
$\delta X^\Phi = 0$. Thus, the perturbation of the fluid is completely 
characterized by its radial ``Lagrangian displacement''
\begin{equation}
\xi(r,t) \equiv \delta X^R (r,t) \, .
\end{equation}
which describes the radial displacement of each fluid element from its
``equilibrium position''. 

Given our choice of $X^A$, in order to be compatible with our
assumption of spherical symmetry, the three-form $N_{ABC}$ on ``fluid space''
must take the form
\begin{equation}
N_{R \Theta \Phi} = q(X^R) \sin X^\Theta,
\end{equation}
for some function $q$.  By \eqref{nuandN}, the number density,
$\nu$, of the fluid is then given by
\begin{equation}
\label{nuandh}
\nu = \frac{q(X^R)}{r^2} \sqrt{ e^{-2 \Lambda} \left(\frac{\partial
  X^R}{\partial r}\right)^2 - e^{-2 \Phi}
  \left( \frac{\partial X^R}{\partial t}\right)^2 }.
\end{equation}
The variation of this formula yields
\begin{equation}
  \label{chandraCrreq}
  \delta \nu = \nu \left( \frac{\partial \xi}{\partial r} + \left(
  \frac{1}{\nu} \frac{\partial 
  \nu}{\partial r} + \frac{\partial \Lambda}{\partial r} + \frac{2}{r}
  \right) \xi - \lambda \right). 
\end{equation}
This formula will enable us to express all terms in the linearized
equations of motion that arise from the variation of $\nu$ in terms of
our chosen dynamical variable $\xi$.

Our next step is to obtain an expression for $F$ and solve the equation
$F=0$. Since, in this case, our matter fields $X^A$ are scalars, the 
terms in \eqref{vdef}
that depend on the matter equations of motion vanish, and we simply have
\begin{equation}
  V_{abc} = 2 \epsilon^d {}_{abc} t^e (\mathcal{E}_G)_{de}.
\end{equation}
From this, we can conclude that
\begin{align}
  V_{t \theta \varphi} &= - 2 \sqrt{-g} (\mathcal{E}_G)_{tr} g^{rr}, &
  V_{r \theta \varphi} &= 2 \sqrt{-g} (\mathcal{E}_G)_{tt} g^{tt}.  
\end{align}
Since $\delta \form{V} = t^a \delta \form{\mathcal{C}}_{a}$, we can therefore
identify the quantity $H_1$ defined in \eqref{constraintform} as
\begin{equation}
  H_1 (t,r) = -2 r^2 e^{\Phi - \Lambda} (\delta \mathcal{E}_G)_{tr},
\end{equation}
Calculating the first-order equation of motion
$(\delta \mathcal{E}_G)_{tr}$, we find that
\begin{equation}
(\delta \mathcal{E}_G)_{tr} = \frac{2}{r} \frac{\partial \lambda}{\partial t} 
- e^{2 \Lambda} \varrho' \nu \frac{\partial \xi}{\partial t} \, .
\end{equation}
From \eqref{constraintform2}, we can immediately deduce that 
\begin{equation}
  \label{chandraconstr}
  F(t,r) = 2 r^2 e^{\Phi - \Lambda} \left( \varrho' \nu \xi - \frac{2}{r} 
  e^{- 2 \Lambda} \lambda \right) = 0.
\end{equation}
Thus, $\lambda$ can be eliminated in terms of the matter variable $\xi$ by
\begin{equation}
  \label{lambdasol}
\lambda = \frac{r}{2} e^{2 \Lambda} \varrho' \nu \xi \, .
\end{equation}

The next step is to write down and solve the equation $(\delta C)_{rr} = 0$
for $\partial \phi / \partial r$. By direct calculation, we obtain
\begin{multline}
(\delta C)_{rr} = (\delta \mathcal{E}_G)_{rr} \\ = \frac{2}{r} e^{- 2
    \Lambda} \left( \frac{\partial \phi}{\partial r} - \left( 2
    \frac{\partial \Phi}{\partial r} + \frac{1}{r} \right) \lambda
    \right) - \varrho'' \nu \delta \nu, 
\end{multline}
where $\delta \nu$ is given by \eqref{chandraCrreq}.
Thus, the solution to $(\delta C)_{rr} = 0$ is\footnote{In writing the equation
in this form, we have used the background equation of motion
$\frac{2}{r} \left( \frac{\partial \Phi}{\partial r} + \frac{\partial
  \Lambda}{\partial r} \right) = e^{2 \Lambda} \varrho' \nu$.} 
\begin{multline}
  \label{phisol}
\frac{\partial \phi}{\partial r} = \left( \frac{\partial \Phi}{\partial r} + 
\frac{\partial \Lambda}{\partial r} \right) \left( 2\frac{ \partial
  \Phi}{\partial r} +  
\frac{1}{r} \right) \xi \\ + \frac{r}{2} e^{2 \Lambda} \varrho'' \nu^2
\left( \xi' +  
\left( - \frac{\partial \Phi}{\partial r} + \frac{2}{r} + \frac{1}{\nu} \frac{ 
\partial \nu}{\partial r} \right) \xi \right).
\end{multline}

Next, we need to write down the linearized matter equations of motion and
substitute our solutions, \eqref{lambdasol} and \eqref{phisol},
for $\lambda$ and $\partial \phi / \partial r$ into these
equations to obtain equations written purely in terms of the 
perturbed matter variables.  In our case, the linearization of
\eqref{euler} yields only one non-trivial equation for our single
nontrivial perturbed matter variable $\xi$. By direct calculation, we
obtain
\begin{widetext}
\begin{equation}
\varrho' \left[ -e^{2 \Lambda - 2 \Phi} \frac{\partial^2 \xi}{\partial
    t^2} + \frac{\partial \phi}{\partial r} 
    \right] \\ + \left( \frac{\partial}{\partial r} + \frac{\partial
    \Phi}{\partial r} \right)  
    \left[ \varrho'' \nu \left(
    \frac{\partial \xi}{\partial r} + \left( \frac{1}{\nu}
    \frac{\partial \nu}{\partial r} + \frac{\partial \Lambda}{\partial
    r} + \frac{2}{r} \right) \xi - \lambda \right)\right] = 0.
\end{equation}
\end{widetext}
Substituting our solutions for $\lambda$ and $\partial \phi / \partial r$,
we obtain
\begin{multline}
\varrho' \left[ -e^{2 \Lambda - 2 \Phi} \frac{\partial^2 \xi}{\partial
    t^2} + \left( 2 \frac{\partial \Phi}{\partial r} + \frac{1}{r}  
  \right) \left( \frac{\partial \Phi}{\partial r} + \frac{\partial
    \Lambda}{\partial r} \right) \xi \right] 
\\ + e^{-2 \Phi - \Lambda} \frac{\partial}{\partial r} 
    \left[ \frac{e^{3 \Phi + \Lambda}}{r^2} \varrho''
    \frac{\partial}{\partial r}  
    \left( r^2 e^{- \Phi} \nu \xi \right) \right] = 0.
\end{multline}
Fortunately, this equation takes the form of \eqref{simpleform}, and it
is straightforward to read off the operator $\mathcal T$ from this equation.

The next step is to evaluate the pullback of the symplectic
current---or, equivalently, $t_a \omega^a$---so as to obtain the inner
product that will appear in the variational principle. As previously
noted $\omega^a$ consists of two pieces, $\omega^a =
\omega^a_\text{grav} + \omega^a_\text{mat}$, given by
\eqref{omegagrav} and \eqref{omegamat} respectively.  By direct
evaluation, we find that for spherically symmetric perturbations of a
static, spherically symmetric background, we have $t_a
\omega^a_\text{grav} = 0$, i.e., there is no ``gravitational
contribution'' to the symplectic form\footnote{This fact is
undoubtedly directly related to the fact that there are no dynamical
degrees of freedom of the gravitational field in the spherically
symmetric case.}. In a static background, the expression for $t_a
\omega^a_\text{mat}$
simplifies considerably, since the fluid 
four-velocity $U^a$ will
be parallel to $t^a$, so $t_a N^{abc} = \nu t_a U_d
\epsilon^{abcd} = 0$. We obtain
\begin{multline}
\label{csympcurr}
t_a \omega^a_\text{mat} = - t_a \frac{\varrho'}{2 \nu} N_{ABC} \nabla_b
X^B \nabla_c X^C \delta_2 X^A \left[ \delta_1 g^{ad} N_{d} {}^{bc}
\right. \\ \left. + 3
\nabla^{[a} (N_{DEF} \delta_1 X^D) \nabla^b X^E \nabla^{c]} X^F \right] - [
  1 \switch 2 ].
\end{multline}
The first term in the square brackets also vanishes in the spherically
symmetric case. Thus, we find that for spherically symmetric 
perturbations of static, spherically symmetric backgrounds,
the symplectic form
\begin{equation}
\Omega = \int_\Sigma \form{\bar{\omega}} = -
\int_\Sigma \dif^3 x \sqrt{h} (\omega^a_\text{grav} + 
\omega^a_\text{mat} ) n_a,
\end{equation}
(where $\Sigma$ is a static slice in the background spacetime, $n^a$ is its
future-directed unit normal, and $\sqrt{h}$ is the volume element
associated with the induced Riemannian metric $h_{ab}$ on $\Sigma$) is
given by 
\begin{multline}
  \Omega =  \frac{3}{2} \int \dif^3 x \, \sqrt{h} n_a 
  \frac{\varrho'}{\nu} \left(N_{ABC}  \delta_2 X^A \nabla_b  X^B \nabla_c 
  X^C \right) \\
  \times \left( N_{DEF} \nabla^{[a} \delta_1 X^D \nabla^b X^E
    \nabla^{c]} X^F \right)  
  - [ 1 \switch 2 ].
\end{multline}
In writing the above, we have used the fact that
there is only one non-vanishing component of $\delta X^A$, so any
term proportional to $\delta_1 X^A \delta_2 X^B$ (as opposed to terms
depending on the derivatives of $\delta X^A$) will vanish under
antisymmetrization.  Writing this out in terms of our perturbational
variables yields the simple expression 
\begin{equation}
\label{xisympform}
\Omega = 4 \pi \int \dif r \, r^2 e^{3 \Lambda - \Phi} 
\varrho' \nu \left( \frac{\partial \xi_1}{\partial t} \xi_2 - \xi_1
\frac{\partial \xi_2}{\partial t} \right). 
\end{equation}
The inner product to be used in our variational principle can now be
read off by comparing \eqref{xisympform} with \eqref{innerprod}. We obtain
\begin{equation}
\label{chandrainnerprod}
(\xi_1, \xi_2) = 4 \pi \int \dif r \, r^2 e^{3 \Lambda - \Phi} 
\varrho' \nu \xi_1 \xi_2
\end{equation}
Since $\varrho' \nu = \rho + P$ under the substitutions in
\eqref{rhoandP}, this quadratic form will be positive for any fluid
satisfying the null energy condition (i.e., $ \rho + P> 0$.)

Our variational principle is then of the form \eqref{simplevarprin}, with 
numerator
\begin{widetext}
\begin{equation}
\label{varprinnumer}
( \xi, \mathcal{T} \xi) = 12 \pi \int \dif r \, \left[ - r^2
  e^{\Lambda  
    + \Phi} \varrho' \nu \left( 2 \frac{\partial \Phi}{\partial r}
  + \frac{1}{r} \right) \left( \frac{\partial \Phi}{\partial r}
  + \frac{\partial \Lambda}{\partial r} \right) \xi^2 \right. 
  \\ \left. + \frac{e^{3 \Phi + \Lambda}}{r^2} \varrho'' \left(
  \frac{\partial}{\partial r}  
  \left( r^2 e^{- \Phi} \nu \xi \right) \right)^2 \right]
\end{equation}
\end{widetext}
and denominator 
\begin{equation}
\label{varprindenom}
(\xi, \xi) = 12 \pi \int \dif r \, r^2 e^{3 \Lambda - \Phi} 
\varrho' \nu \xi^2
\end{equation}
This is equivalent to the variational principle originally derived by
Chandrasekhar\footnote{Note that Chandrasekhar's $\gamma$ can be
obtained by the substitution $\varrho'' \nu^2 \to \gamma P$ in the
Lagrangian coordinate formalism.} \cite{Chandra1, Chandra2}, but we
have rederived it here in a systematic way that is essentially
``foolproof''. 

\section{Conclusion}

We have presented a general procedure to analyse the stability of
spherically symmetric perturbations about static spherically symmetric
solutions of an arbitrary covariant field theory.  This procedure
involves solving the linearized constraints and eliminating the metric
perturbation variables algebraically.  The symplectic form is then
used to define an inner product, from which a variational principle
can be obtained.

It is important to emphasize that our procedure is entirely prescriptive, as
illustrated by the outline in Section \ref{Chandra};  while the method
can fail at various points (e.g., the inner product obtained from the
symplectic form could fail to be positive), there is in principle
never any ``art'' involved in applying this procedure to a given field
theory.  This method is therefore a potentially powerful tool for
analyzing the viability alternative theories of gravity and other
covariant field theories; in an upcoming work \cite{VPA}, we will
apply this method to three alternative theories of gravity \cite{CDTT,
  Aether1, TeVes}. 

\begin{acknowledgments}
This research was supported by NSF grant PHY-0456619 to the University
of Chicago and is based on work supported under a National Science
Foundation Graduate Research Fellowship to M.S. M.S.\ also gratefully
acknowledges the support of the ARCS Foundation of Chicago.
\end{acknowledgments}


\end{document}